\documentclass[12pt]{article}
\usepackage[english,german,french,polish]{babel}
\usepackage[T1]{fontenc}
\usepackage{amsfonts}

\begin{document}

\baselineskip 0.75cm
\topmargin -0.6in
\oddsidemargin -0.1in

\let\ni=\noindent

\renewcommand{\thefootnote}{\fnsymbol{footnote}}

\newcommand{\SM}{Standard Model }

\newcommand{\SMo}{Standard-Model }

\pagestyle {plain}

\setcounter{page}{1}

\pagestyle{empty}

~~~

\begin{flushright}
IFT-- 10/15
\end{flushright}

\vspace{0.3cm}

{\large\centerline{\bf Predictive empirical mass formula}}
{\large\centerline{\bf for up and down quarks of three generations{\footnote{Work supported in part by Polish MNiSzW scientific research grant N N202 103838 (2010--2012).}} }}

\vspace{0.5cm}

{\centerline {\sc Wojciech Kr\'{o}likowski}}

\vspace{0.3cm}

{\centerline {\it Institute of Theoretical Physics, University of Warsaw }}

{\centerline {\it Ho\.{z}a 69,~~PL--00--681 Warszawa, ~Poland}}

\vspace{1.2cm}

{\centerline{\bf Abstract}}

\vspace{0.3cm}

The starting point in this note is the charged-lepton mass formula successfully predicting the figure $m_\tau = 1776.80$ MeV for $\tau$-lepton mass {\it versus} its recent experimental value $m_\tau = 1776.82 \pm 0.16$ MeV. There, two experimental masses $m_e$ and $m_\mu$ are the only input. This two-parameter empirical mass formula is extended to a pair of three-parameter empirical mass formulae for up and down quarks of three generations where, subsequently, a third pair of its free parameters is conjectured to be specified in a suggestive way, being replaced by one free parameter. Then, the new quark mass formulae predict one quark  mass $m_s = 101$ MeV {\it versus} its experimental value $m_s = 101^{+29}_{-21}$ MeV. Here, five experimental masses $m_u , m_c , m_t$ and $m_d , m_b$ (their central values, for simplicity) are the only input.

\vspace{0.6cm}

\ni PACS numbers: 12.15.Ff , 12.90.+b .

\vspace{0.8cm}

\ni November 2010

\vfill\eject

~~~
\pagestyle {plain}

\setcounter{page}{1}

\vspace{0.2cm}

\ni {\bf 1.Introduction: charged leptons}

\vspace{0.2cm}

In this note, we try to extend to quarks the charged-lepton mass formula proposed by us in 1992 [1,2],

\begin{equation}
m_N  =  \rho_N \,\mu\! \left(\!N^2 + \frac{\varepsilon -1}{N^2}\! \right) \,,
\end{equation} 

\ni where $N = 1,3,5$ is a quantum number, $\mu$ and $\mu(\varepsilon -1)$ denote two mass-dimensional free parameters, while


\begin{equation}
m_e  \equiv m_1\;,\;  m_\mu  \equiv m_3\;,\;  m_\tau  \equiv m_5
\end{equation}

\ni are charged-lepton masses and


\begin{equation} 
\rho_1 = \frac{1}{29} \;,\; \rho_3 = \frac{4}{29} \;,\; \rho_5 = \frac{24}{29} 
\end{equation}

\ni ($\sum_N\rho_N = 1$) stand for some fermion generation-weighting factors. The explicit form of the mass formula (1) is

\begin{eqnarray}
m_e & = & \frac{\mu}{29} \,\varepsilon  \,, \nonumber \\
m_\mu & = & \frac{4\mu}{29}\,\frac{80 +\varepsilon}{9}  \,, \nonumber \\
m_\tau & = & \frac{24\mu}{29}\,\frac{624 + \varepsilon}{25} \,. 
\end{eqnarray}

Eliminating from Eqs. (4) two free parametrs $\mu$ and $\varepsilon$, we obtain the mass sum rule [1,2],

\begin{equation}
m_\tau = \frac{6}{125} \left(351m_\mu - 136 m_e\right) = 1776.80\;{\rm MeV} 
\end{equation}

\ni (or, more precisely, $m_\tau = 1776. 7964$ MeV), and determine both free parameters,

\begin{equation}
\mu = \frac{29 (9m_\mu - 4 m_e)}{320} = 85.9924\;{\rm MeV} \;\;,\;\;\; \varepsilon = \frac{320 m_e}{9 m_\mu - 4 m_e} = 0.172329 \,,
\end{equation}

\ni when we use the experimental values $m_e = 0.5109989$ MeV and $m_\mu = 105.65837$ MeV as the only input. We can see that the value $m_\tau = 1776.80$ MeV (or, more precisely, $m_\tau = 1776.7964$ MeV) calculated in Eq. (5) is our prediction for $\tau$-lepton mass [1,2].

It happens that for several years the measured value of the $\tau$-lepton mass continues to approach closer and closer the particular figure $m_\tau = 1776.80$ MeV, namely


\begin{eqnarray} 
m^{(2004)}_{\tau} & = & 1776.99^{+0.29}_{-0.26}\;{\rm MeV} \;, \\
m^{(2006)}_{\tau} & = & 1776.99^{+0.29}_{-0.26}\;{\rm MeV} \;, \\
m^{(2008)}_{\tau} & = & 1776.84 \pm 0.17\;{\rm MeV} \,,\\
m^{(2010)}_{\tau} & = & 1776.82 \pm 0.16\;{\rm MeV} 
\end{eqnarray} 

\ni ({\it cf.} Refs.  [3], [4], [5], [6], respectively). Such an exciting convergence on the value (5) of $m_\tau$ is very promising for our mass formula (1).{\footnote{ In 1981, Koide proposed for charged-lepton masses another approach based on the neatly looking nonlinear equation symmetrical in $m_e$, $m_\mu$ and $m_\tau$ [7],
$$
m_e + m_\mu + m_\tau = \frac{2}{3}(\sqrt{m_e} + \sqrt{m_\mu} + \sqrt{m_\tau})^2,
$$
giving two solutions for $m_\tau$ in terms of experimental values of $m_e$ and $m_\mu$:
$$
m_\tau = \left[2(\sqrt{m_e} + \sqrt{m_\mu} )\pm \sqrt{3(m_e + m_\mu)+12\sqrt{m_e m_\mu}}\, \right]^2 = \left\{\begin{array}{r} 1776.97\,{\rm MeV} \\ 3.31735\,{\rm MeV}  \end{array} \right.\,,
$$ 
where $m_e$ = 0.5109989 MeV and $m_\mu$ = 105.65837 MeV. The first solution still agrees wonderfully with the central value of actual experimental estimate (10), though its small deviation from this experimental value is  larger than the tiny deviation of our prediction (5). The second solution gets no interpretation yet.}}

In this mass formula, the charged-lepton mass may be interpreted [1,2] as a  linear combination of two intrinsic couplings within a structure of pointlike lepton built up of the quantum number $N = 1,3,5$ (the only one being here at our disposal) equal to the number of "intrinsic partons"~participating in three fermion generations, where each "intrinsic parton"~is described by a Dirac bispinor index at the wave function. These two couplings are [1,2]:

\begin{description}
\item(i) $N^2 = N + N(N-1)$ --- intrinsic self-coupling plus mutual coupling of all "intrinsic partons"~treated on equal footing, and

\item(ii) $N^{-2} = [N!/(N-1)!]^{-2}$ --- a correction to intrinsic self-coupling of an "intrinsic parton"~distinguished from the other $N-1$ "intrinsic partons"~by its presumed correlation with the set of \SMo labels, while these other $N-1$ "intrinsic partons"~are indistinguishable, what gives $ [N!/(N-1)!]^{-1}$ as the appearance probability of the distinguished "intrinsic parton".
\end{description}

In consequence of this interpretation, $N-1$ indistinguishable "intrinsic partons"~obey the Fermi statistics along with the "intrinsic Pauli principle"~which requires $N-1$ Dirac bispinor indices at the wave function to be antisymmetrized, when they describe these $N-1$ "intrinsic partons". This implies that $N-1 \leq 4$ and so,  that $N= 1.3.5$ {\it i.e.}, there are three fermion generations (because each Dirac bispinor index can assume four different values 1,2,3,4). Of course, at the wave function there can be $N=1,3,5$ Dirac bispinor indices, but among them only one is presumed to be distinguished as being correlated with the set of \SMo labels. In this interpretation, the particular values (3) for fermion-generation weighting factors $\rho_N$ follow naturally [1,2].

\vspace{0.2cm}

\ni {\bf 2. Up and down quarks}

\vspace{0.2cm}

Unfortunately, the mass formula of the shape (1) with two pairs of free parameters $\mu^{(u,d)}$ and $\varepsilon^{(u,d)}$ cannot be fitted successfully to the observed mass spectrum of up and down quarks of three generations. In order to extend such a formula, we may introduce into up and down linear combinations a third pair of free parameters $\eta^{(u,d)}$ multiplied by the third intrinsic coupling:

\begin{description}
\item(iii) $-(N-1)$ --- a corrective intrinsic coupling of $N-1$ indistinguishable "intrinsic partons"~within the quark as a whole.
\end{description}

Such an extension leads to the quark mass formula of the shape

\vspace{-0.1cm}

\begin{equation}
m_N^{(u,d)}  =  \rho_N \mu^{(u,d)} \left[ N^2 + \frac{\varepsilon^{(u,d)}-1}{N^2}- \eta^{(u,d)}(N-1)\right] 
\end{equation}

\ni with $N = 1,3,5$. Explicitly, the mass formula (11) gives 

\vspace{-0.2cm}
 
\begin{eqnarray}
m_{u,d} & = &  \frac{\mu^{(u,d)}}{29}\varepsilon^{(u,d)}  \,,\nonumber \\
m_{c,s} & = &  \frac{4\mu^{(u,d)}}{29}\; \frac{80+\varepsilon^{(u,d)} - 18\eta^{(u,d)}}{9} \,,\nonumber \\
m_{t,b} & = &  \frac{24\mu^{(u,d)}}{29}\; \frac{624+\varepsilon^{(u,d)} - 100\eta^{(u,d)}}{25} \,,
\end{eqnarray} 

\vspace{-0.1cm}

\ni where

\vspace{-0.2cm}

\begin{equation}
m_{u,d} \equiv  m^{(u,d)}_1 \;,\; m_{c,s} \equiv  m^{(u,d)}_3 \;,\; m_{t,b} \equiv  m^{(u,d)}_5 \,.
\end{equation}

\ni Here, there are six free parameters $\mu^{(u,d)}\,,\,\varepsilon^{(u,d)}$ and $\eta^{(u,d)}$ that enable us to fit successfully our mass formula to the observed quark spectrum [6]:

\begin{equation}
m_{u,d} \!=\!\! \left\{ \begin{array}{rrrr}\!\!1.7\!\! & \!\!{\rm to}\!\! & \!\!3.3\!\! & \!\!{\rm MeV}\!\!\\ \!\! 4.1\!\! & \!\!{\rm to}\!\! & \!\!5.8\!\! & \!\!{\rm MeV}\!\! \end{array}\right. \!\!\rightarrow\!\! \left\{ \begin{array}{rr}\!\!2.5\!\! & \!\!{\rm MeV}\!\! \\5.0 \!\! & \!\!{\rm MeV}\!\! \end{array}\right.\,,\, 
m_{c,s} \!=\!\! \left\{\begin{array}{lr}\!\!1.27^{+0.07}_{-0.09}\!\! & \!\!{\rm  GeV}\!\! \\ \!\!101^{+29}_{-21}\!\!\!\!&\!\!\!\!{\rm MeV}\!\!  \end{array}\right.\,,\, 
m_{t,b} \!=\!\! \left\{\begin{array}{l}\!\!172.0\pm 2.2\;\, {\rm GeV}\!\! \\ \!\!4.19^{+0.18}_{-0.06}\;\,{\rm GeV}\!\! \end{array}\right.
\end{equation}

\ni (for $m_{u,d}$ we take mean values).

If six free parameters are really independent, our quark mass formula (11) or (12) gives no mass predictions, only determining six free parameters in terms of six quark masses through the following inverse relations to Eqs. (12):

\begin{eqnarray}
\mu^{(u,d)} & = &  \frac{29}{12928}\left[75m_{t,b}-4(225m_{c,s} - 82m_{u,d})\right] =  \left\{\begin{array}{lr}\!\!26.4\!\!& {\rm GeV}\!\! \\\!\!~\,0.505\!\! & {\rm GeV}\!\!  \end{array}\right.\;, \\  & & \nonumber \\
\varepsilon^{(u,d)} & = &  \frac{29}{\mu^{(u,d)}} m_{u,d} = \frac{12928 m_{u,d}}{75m_{t,b}-4(225m_{c,s} - 82m_{u,d})} = \left\{\begin{array}{l}\!\!0.0027\!\! \\\!\!0.29\!\!\end{array}\right.\;, \\ & & \nonumber \\
\eta^{(u,d)} & = &  \frac{8}{3}  \frac{125 m_{t,b}-6(351m_{c,s} - 136m_{u,d})}{75m_{t,b}-4(225m_{c,s} - 82m_{u,d})} = \left\{\begin{array}{ll}\!\!4.27\!\! & = 4 + 0.27 \\\!\!3.73\!\! & = 4 - 0.27\end{array}\right.\;. 
\end{eqnarray} 

\ni Defining two new free parameters

\begin{equation}
\pm \omega^{(u,d)} \equiv \eta^{(u,d)}  - 4 
\end{equation}

\ni dependent on

\begin{equation}
\eta^{(u,d)} \equiv  4 \pm \omega^{(u,d)} \,,
\end{equation}

\ni we find from Eqs. (17) the relations

\begin{equation}
\omega^{(u,d)} = \pm \frac{4}{3}\; \frac{25m_{t,b}-216(7m_{c,s} - 3m_{u,d})}{75m_{t,b}-4(225m_{c,s} - 82m_{u,d})}  = \left\{\begin{array}{l}\!\!0.270\!\! \\\!\!0.265 \end{array}\right.\;. 
\end{equation}

\vspace{0.2cm}

\ni In Eqs. (15), (16), (17) and (20), all six experimental masses (14) (their central values, for simplicity) are used as input to determine all six free parameters. Note that the numerator in Eq. (17) has the form of a linear combination of three masses like in the charged-lepton mass sum rule (5). Thus, if Eq. (5) holds, then the correspondingly defined $\eta$ is zero for charged leptons as it ought to be.

Now, we make the conjecture suggested by Eqs. (20) (implying approximate equality of  $\omega^{(u)}$ and $\omega^{(d)}$) that, in reality, we have strictly

\begin{equation}
\omega^{(u)} = \omega^{(d)}\;\;,\;\; {\rm with}\;\; \omega \equiv \omega^{(u)} = 0.270 \,,
\end{equation}

\ni where experimental masses $m_u , m_c , m_t$ are taken as an input ($m_c$ and $m_t$ being better known experimentally than $m_s$ and $m_b$). Then, the former free parameters $\omega^{(u)}$ and $\omega^{(d)}$ (and so, the dependent $\eta^{(u)}$ and $\eta^{(d)}$) are specified in a suggestive way, replaced by one free parameter $\omega \equiv \omega^{(u)}$. In this case, from Eqs. (20) and (21), we derive the following predictive quark mass sum rules:

\begin{equation}
25(4 \mp 9\omega)m_{t,b} - 108 (56 \mp 25\omega) m_{c,s} + 24(108 \mp 41\omega) m_{u,d} = 0\,,  
\end{equation}

\ni where $\omega$ is a free parameter assuming the value (21) with the input of experimental masses $m_u , m_c , m_t$.

If we decide to treat the mass $m_s$ as unknown (or rather as less known than $m_b$) and all other quark masses as experimentally given, we get from Eqs. (22) the mass sum rule predicting

\begin{equation}
m_s  =  \frac{25(4 + 9\omega)m_b+ 24(108 + 41\omega) m_d}{108 (56 + 25\omega)} = 101\;{\rm MeV}
\end{equation}

\ni {\it versus} the experimental value $m_s = 101^{+29}_{-21}$ MeV. In Eq. (23), five experimental quark masses (their central values, for simplicity) among six are applied as the only input: first, $m_u , m_c , m_t$ are used to calculate $\omega \equiv \omega^{(u)}$ in Eq. (20) and then, $m_d , m_b$ to evaluate $m_s$ from Eq. (23) (with the value (21) for $\omega \equiv \omega^{(u)}$). We can see that our prediction (23) for $m_s$ is identical with its experimental central value (14). Due to this, the values (15), (16) and (17) of $\mu^{(d)} , \varepsilon^{(d)}$ and $\eta^{(d)} \equiv 4 - \omega^{(d)}$ determined with the use of experimental masses $m_d , m_s , m_b$ turn out to be selfconsistent with our prediction for $m_s$.

In the case of conjecture (21) for the parameters (18), our quark mass formula (11) and its explicit form (12) become  

\begin{equation}
m^{(u,d)}_N  =  \rho_N \;\mu^{(u,d)} \left[N^2 + \frac{\varepsilon^{(u,d)}-1}{N^2} - (4\pm \omega)(N-1) \right] 
\end{equation}

\ni with $N = 1,3,5$, and

\vspace{-0.4cm}

\begin{eqnarray}
m_{u,d} & = & \frac{\mu^{(u,d)}}{29} \varepsilon^{(u,d)} \,, \nonumber \\
m_{c,s} & = & \frac{4\mu^{(u,d)}}{29}\, \frac{8+\varepsilon^{(u,d)} \mp 18\,\omega}{9} \,, \nonumber \\
m_{t,b} & = & \frac{24\mu^{(u,d)}}{29}\, \frac{224+\varepsilon^{(u,d)} \mp 100\,\omega}{25}  \,, 
\end{eqnarray}

\vspace{-0.1cm}

\ni respectively, implying predictive mass sum rules (22).

\vspace{0.2cm}

\ni {\bf 3. Conclusion}

\vspace{0.2cm}

In this note, we presented two empirical mas formulae: for charged leptons, Eq. (1), involving two free parameters $\mu$ and $\varepsilon$, and for up and down quarks, Eqs. (11), depending primarily on six free parameters, $\mu^{(u,d)} , \varepsilon^{(u,d)}$ and $\eta^{(u,d)} \equiv 4 \pm \omega^{(u,d)}$, and eventually on five ones,  $\mu^{(u,d)} , \varepsilon^{(u,d)}$ and $\omega \equiv \omega^{(u)}$, if it is guessed that  $\omega^{(u)} = \omega^{(d)}$. Then, with five other experimental quark masses taken as the only input, our prediction is $m_s = 101$ MeV.

The charged-lepton mass formula, being in an excellent agreement with the experimental data, was our starting point to an extension leading to the quark mass formulae which turned out to be consistent with the experiment as well.

In the past, we proposed for up and down quarks of three generations some other well working mass formulae [8] depending also on three pairs of free parameters, but then we were not able to reduce the number of their six free parameters in a satisfactory way (with $m_{u,d} \neq 0$) in order to obtain predictive mass sum rules. 

From the viewpoint of the conventional quantum field theory (in contrast to quantum mechanics of composite systems) the dependence of a fundamental-particle mass spectrum on discrete quantum numbers (as our $N = 1,3,5$) may seem nonrealistic. We believe, however, that our intrinsic composite interpretation of leptons and quarks [1.2] introduces a needed quantum-mechanical aspect into particle physics and so, discrete mass formulae can work for fundamental particles.

Finally, we would like to stress an obvious critical point that appears in our approach, when we accept seriously the particular quark-mass estimates (14), though -- in contrast to charged leptons -- quarks are not asymptotic states of \SM and their masses depend in principle on renormalization procedure. Nevertheless, if these  mass estimates are somehow selected to hold as in Ref. [6], our approach to such quark masses works and is consistent with the conjecture of $\omega^{(u)} = \omega^{(d)}$. 

\vfill\eject

~~~~
\vspace{0.5cm}

{\centerline{\bf References}}

\vspace{0.5cm}

{\everypar={\hangindent=0.6truecm}
\parindent=0pt\frenchspacing

{\everypar={\hangindent=0.6truecm}
\parindent=0pt\frenchspacing

~[1]~For a recent presentation {\it cf.} W.~Kr\'{o}likowski, {\it Acta Phys. Polon.}, {\bf B 41}, 649 (2010); and references therein. 

\vspace{0.2cm}

~[2]~For a comparison with the 2010-experimental value of $m_\tau$ {\it cf.} W.~Kr\'{o}likowski,  arXiv: 1009.2388 [{\tt hep--ph}] (unpublished).

\vspace{0.2cm}

~[3]~S.~Eidelman {\it et al.} ( Particle Data Group), {\it Phys. Lett.}, {\bf B 592}, 1 (2004).

\vspace{0.2cm}

~[4]~W.M.~Yao {\it et al.} ( Particle Data Group), {\it J. Phys}, {\bf G 33}, 1 (2006).

\vspace{0.2cm}

~[5]~C.~Amsler {\it et al.} (Particle Data Group), {\it Phys. Lett.} {\bf B 667}, 1 (2008).

\vspace{0.2cm}

~[6]~N.~Nakamura {\it et al.} (Particle Data Group), {\it J. Phys}, {\bf G 37}, 075021 (2010).

\vspace{0.2cm}

~[7]~For a more recent discussion {\it cf.} Y.~Koide, {\tt hep--ph/0506247}; and references therein.

\vspace{0.2cm}

~[8]~W. Kr\'{o}likowski, {\it Acta Phys. Polon.} {\bf B 37}, 2601 (2006).

\vspace{0.2cm}

\vfill\eject

\end{document}